\documentstyle[12pt,psfig]{article}
\newcommand{\mathbold}[1]{\mbox{\protect\boldmath $\displaystyle #1$}}
\begin{document}
\title{$J/\Psi$ production in ${\pi}N$ collisions
\footnote{Supported 
by Forschungszentrum J\"ulich and the Australian Research Council}}
\author{
A. Sibirtsev$^1$~\footnote{sibirt@theo.physik.uni-giessen.de,
on leave from the Institute of Theoretical and Experimental Physics,
117259 Moscow, Russia} , \
K. Tsushima$^2$~\footnote{ktsushim@physics.adelaide.edu.au} \\
{$^1$\small Institut f\"ur Theoretische Physik, Universit\"at Giessen}\\
{\small D-35392 Giessen, Germany} \vspace{0.5em}\\
{\small $^2$Center for the Subatomic Structure of Matter (CSSM)}\\
{\small and Department of Physics and Mathematical Physics}\\
{\small The University of Adelaide, SA 5005, Australia}}
\date{ }
\maketitle

\begin{abstract}
We calculate the total cross section for 
$\pi N \to J/\Psi N$ based on $\rho$-meson exchange.
On the basis of this calculation we predict a maximal reaction 
rate arising from the OZI-violating $J/\Psi{\leftrightarrow}\rho\pi$ 
vertex in tree-level. 
Our estimate for the maximum reaction rate is still one 
order of magnitude smaller than the existing predictions 
which were based on the other OZI-violating mechanisms.
\end{abstract}
\vspace{1cm}

%
\vspace{-12.8cm}
\hfill UGI-98-32, ADP-98-63/T330
\vspace{12.8cm}
%

The production of $J/\Psi$-mesons in hadronic reactions is 
an extremely sensitive tool for studying the Okubo-Zweig-Iizuka 
(OZI) rule~\cite{OZI}. The OZI rule is based on the  
assumption that the contribution from processes  
involving the production of quark-antiquark ($q\bar{q}$) 
pairs not available in the initial state ideally equals zero. 
Thus, a violation of the OZI rule might be regarded as evidence for a 
hidden $q\bar{q}$ component in the initial hadrons~\cite{Ellis}. 
In case of the $J/\Psi$ production in ${\pi}N$ collisions 
the OZI violation directly leads to the percentage of 
the $c\bar{c}$ component in the nucleon, or hidden charm~\cite{Chiang}.

There are several studies of the $J/\Psi$ production based on the 
OZI-violating mechanisms. For example, 
within the duality violating model Bolzan et al.~\cite{Bolzan} 
predicted a $\pi^-p{\to}{J/\Psi}n$ total cross section of 1.0 
to 2.5~nb in the threshold enhancement region. A similar result 
was obtaind by Kodaira and Sasaki~\cite{Kodaira} using a generalized 
Veneziano model, which assumed the production of 
$J/\Psi$-meson was through 
mixing with the spin one daughters of the $\omega$ recurrences.
Berger and Sorencsen~\cite{Berger} predicted 
an enhanced rate for the $\pi^-p{\to}{J/\Psi}n$ reaction at 
pion momenta around 10-12~GeV, when the $J/\Psi$-mesons fit 
into a $c\bar{c}$ scheme. The prediction was 
according to the analogy of the study for the 
$\pi^-p{\to}{\phi}n$ reaction, in which a box diagram was adopted 
as the dynamical process that violates the OZI rule.
Okubo~\cite{Okubo1} and Lipkin~\cite{Lipkin} also 
noted that the magnitude of the OZI rule violation depends on 
dynamical mechanisms. 

Here we study the role of the
dynamical input for the OZI-violating, ${\pi}N{\to}{J/\Psi}N$ 
reaction, and calculate the total cross section 
using one of the conventional approaches, $\rho$-meson exchange. 
Note that the vertex, 
$J/\Psi{\leftrightarrow}\rho\pi$, appearing 
in the present treatment is the OZI-violating process in tree-level.

\begin{figure}[t]
\hspace{3cm}
\psfig{file=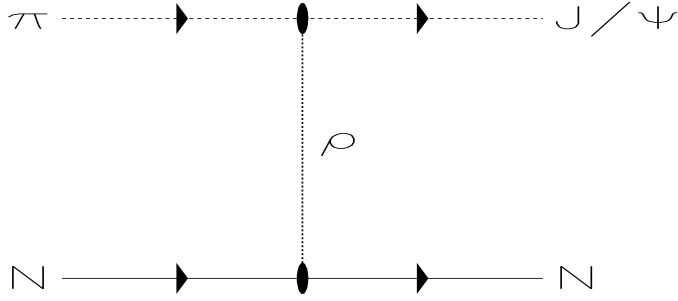,width=10.cm,height=5cm}
\caption[]{Diagram for ${\pi}N{\to}{J/\Psi}N$ reaction
due to the $\rho$-exchange.}
\label{psi1}
\end{figure}

Effective Lagrangian densities relevant for the process depicted in 
Fig.~\ref{psi1} may be given by 
\begin{eqnarray}
{\cal L}_{{\rho}NN} &=& -g_{{\rho}NN}
\left(\bar{N}\gamma^\mu {\mathbold \tau}N \cdot {\mathbold \rho}_\mu
\right. \nonumber \\
 &+& \left. \frac{\kappa}{2m_N} \bar{N} \sigma^{\mu\nu} 
{\mathbold \tau} N \cdot \partial_\mu {\mathbold \rho}_\nu\right),
\label{rnn}\\
{\cal L}_{J\rho\pi} &=& \frac{g_{J\rho\pi}}{m_J} \
\epsilon_{\alpha\beta\mu\nu} ( \partial^\alpha J^\beta )
( \partial^\mu {\mathbold \rho}^\nu ) \cdot {\mathbold \pi}, 
\label{jrp}
\end{eqnarray}
where we use the value, 
$\kappa{\equiv}f_{{\rho}NN}/g_{{\rho}NN}{=}6.1$ and 
$g^2_{{\rho}NN}/4{\pi}{=}0.74$,  
which were used in Ref.~\cite{Tsushima}.

The ${\pi}N{\to}{J/\Psi}N$ amplitude without isospin 
factor is given by:
\begin{eqnarray}
M = \frac{g_{J\rho\pi} g_{{\rho}NN}}{m_J^2} 
\frac{1}{t-m_\rho^2+i\epsilon} \
\epsilon_{\alpha \beta \mu \nu} \ P^\alpha_J \epsilon^{*\beta}_J 
q^\mu  \nonumber \\ \times
\bar{N}(P'_N) \left(\gamma^\nu -\frac{\kappa}{2 m_N}
(P'_N + P_N)^\nu \right) N (P_N), 
\end{eqnarray}
with $\epsilon_J$ being the $J/\Psi$ polarization vector. 

Then the spin-averaged, squared invariant amplitude has the form:
\begin{eqnarray}
|\,\overline{M}\,|^2 
&=& \frac{g^2_{J\rho\pi} \ g^2_{{\rho}NN}}{(t-m_\rho^2)^2}
\ F^2_{J\rho\pi}(t) \  F^2_{{\rho}NN}(t) 
\left\lbrace -2 \, t \, q_{ex}^2 
\right.  \nonumber \\ &+&
\left\lbrack 1-2\kappa + \left. \kappa^2
\left( 1-\frac{t}{4m_N^2} \right) \right\rbrack 
\left\lbrack q_{ex}^2(t-4m_N^2) \right. \right. \nonumber \\
&-& \left. \left. \frac{t}{4m_J^2}\left(2s+t-2m_N^2-
m_J^2-m_\pi^2 \right)^2 
\right\rbrack \right\rbrace, 
\label{A1} 
\end{eqnarray}
where $s$ is the squared invariant collision energy and
\begin{equation}
q_{ex}^2= \frac{(m_J^2-m_\pi^2+t)^2-4m_J^2t}{4m_J^2}.
\end{equation}
The invariant differential cross section is given by 
\begin{equation}
\frac{d\sigma}{dt} = \frac{|\,\overline{M}\,|^2}
{16\pi \ \lambda(s,m_\pi^2,m_N^2)}, 
\end{equation}
with 
\begin{equation}
\lambda (x,y,z) = (x-y-z)^2-4yz.
\end{equation}
The relation between the averaged  
${\pi}N{\to}{J/\Psi}N$  cross section in isospin space 
and different  isospin channels is given by: 
\begin{eqnarray}
\bar{\sigma}({\pi}N{\to}{J/\Psi}N)\!\!&{=}&\!\frac{1}{2} \ 
\sigma(\pi^+n{\to}{J/\Psi}p)
{=}\frac{1}{2} \  \sigma(\pi^-p{\to}{J/\Psi}n) \nonumber \\
&=& \sigma(\pi^0n{\to}{J/\Psi}n) = \sigma(\pi^0p{\to}{J/\Psi}p). 
\end{eqnarray}

Note that above formalism can be used also for vector meson  
production in pseudoscalar +nucleon ($PN{\to}VN$) reactions 
with vector exchange, such as the ${\pi}N{\to}{\omega}N$, 
${\pi}N{\to}{\rho}N$,  ${\pi}N{\to}{\phi}N$ and 
$KN{\to}K^{\ast}N$ reactions, by replacing 
the relevant coupling constants and the form-factors
in Eq.~(\ref{A1}).

For simplicity we use the same form factor
for the vector and tensor couplings in the ${\rho}NN$ vertex:
\begin{equation}
F(t) = \frac{\Lambda^2-m_\rho^2}{\Lambda^2-t}, 
\label{fof}
\end{equation}
with the value for the cut-off parameter, 
$\Lambda{=}920$~MeV~\cite{Tsushima}, 
and $t$ being the 4-momentum transfer from the initial to
the final nucleon.

The $J\rho\pi$ coupling constant, $g_{J\rho\pi}$, 
can be evaluated within the 
narrow resonance approximation for the $\rho$-meson: 
\begin{equation}
\Gamma_{J/\Psi{\to}\rho\pi } = \frac{g^2_{J\rho\pi}}{32\pi}\; 
\frac{\lambda^{3/2}(m_J^2, m_\rho^2, m_\pi^2)}{m_J^5}. 
\label{gjrp1}
\end{equation}

Using the measured partial width, 
$\Gamma_{J/\Psi{\to}\rho{+}\pi}$=31.12 MeV \cite{PDG}, 
we obtain, $g_{J\rho\pi}{=}$ $6.68{\times}10^{-3}$.
Because the $\rho$-meson has a width, the $g_{J\rho\pi}$ 
coupling constant should be determined properly 
in a more rigorous treatment~\cite{Manley}:
\begin{eqnarray}
\Gamma_{J/\Psi{\to}\rho\pi}&=&\frac{g_{J\rho\pi}^2}
{16\pi^2 m_J^5}\!\! \intop_{2m_\pi}^{m_J-m_N} \!\!d\mu \
\lambda^{3/2}(m_J^2,\mu^2,m_\pi^2) \nonumber \\
&\times&
\frac{\mu^2 \ \Gamma_{\rho{\to}2\pi}(\mu)}
{(\mu^2-m_\rho^2)^2+\mu^2\Gamma_{\rho{\to}2\pi}^2(\mu)}, 
\label{gjrp2}
\end{eqnarray}
where the energy dependence of the $\rho$-meson width was taken as 
\begin{equation}
\Gamma_{\rho{\to}2\pi}(\mu)=\Gamma_0 \frac{m_\rho}{\mu} 
\left( \frac{\mu^2-4m_\pi^2}{m_\rho^2-4m_\pi^2} \right)^{3/2}, 
\end{equation}
with $\Gamma_0{=}151.2$~MeV. Obviously, in the limit, 
$\Gamma_{\rho{\to}2\pi}{\to}0$, 
Eq.~(\ref{gjrp2}) approaches the narrow resonance 
approximation of Eq.~(\ref{gjrp1}). According to Eq.~(\ref{gjrp2})
we get the value for the coupling constant, 
$g_{J\rho\pi}{=}6.78{\times}10^{-3}$,
which is close to the result of the narrow 
resonance approximation. This is due to the large upper limit
of the integral in Eq.~(\ref{gjrp2}).

\begin{figure}[t]
\psfig{file=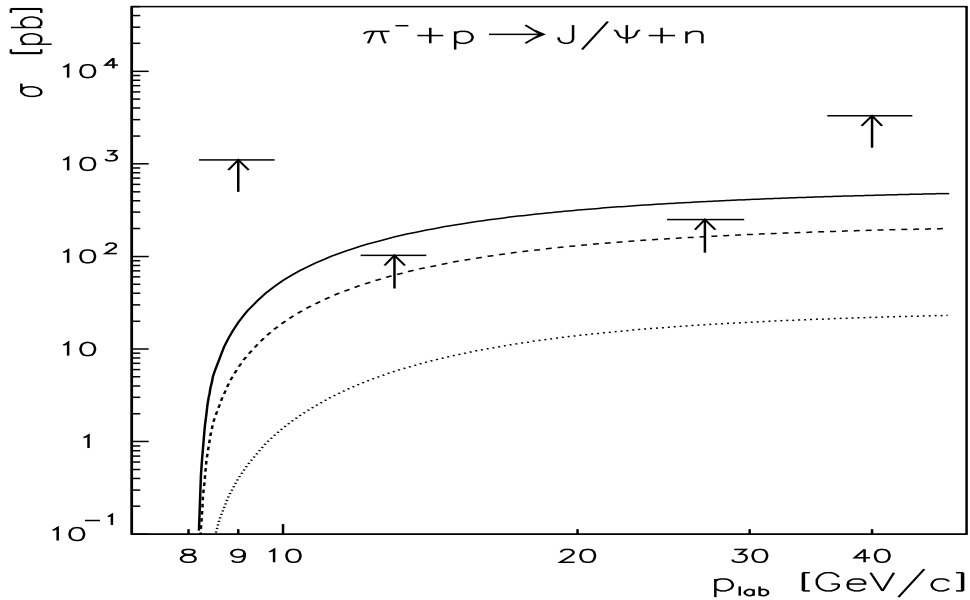,width=15cm,height=9cm}
\caption{The $\pi^-p{\to}{J/\Psi}n$ total cross section. 
Arrows indicate the upper limits evaluated from experimental
data extracted from $\pi A$ 
collisions~\protect\cite{Chiang,Jenkins,Golovkin}. 
Lines show our calculations with the different
cut-off parameters: $\Lambda$=1~GeV (dotted), 
$\Lambda$=2~GeV (dashed) and without form factor
at the $J\rho\pi$ vertex (solid).}
\label{psi2}
\end{figure}

In evaluating the amplitude, one should also 
account for the nonlocality of the $J\rho\pi$ vertex,  
and needs to introduce a relevant form factor. 
To illustrate the sensitivity of the 
calculations to the form factor at the $J\rho\pi$ 
vertex, we use, in the present study, 
the same form factor of Eq.~(\ref{fof})   
and show results for the different  
values of $\Lambda$ in Fig.~\ref{psi2}. Then, our 
estimatate for the maximal reaction rate may correspond to the results
obtained without the form factor at the $J\rho\pi$ vertex. 

Here it should be emphasized that the  
$J/\Psi $ production has never been observed in exclusive 
${\pi}N$ reactions. The upper limits have been set at four
pion beam momenta and actually were evaluated from
${\pi}A$ collisions assuming a linear $A$-dependence 
of the production cross sections~\cite{Chiang,Jenkins,Golovkin}. 

Fig.~\ref{psi2} is the comparison between our results 
and the upper limits for the ${\pi}N{\to}{J/\Psi}N$ total cross 
sections extracted from ${\pi}A$ collisions. 
The dotted, dashed and solid lines show the results with
the values for the cut off parameter, $\Lambda$=1~GeV, 
$\Lambda$=2~GeV, and without the form factor 
at the $J\rho\pi$ vertex, respectively.
The best agreement with the data is for those results obtained without 
the form factor at the $J\rho\pi$ vertex. However, even
this case the predicted maximum cross section is 140~pb,  
at pion laboratory momentum 12~GeV/c.

Although the estimate without the form factor at 
the $J\rho\pi$ vertex can be regarded 
as an upper limit in the present treatment of 
OZI-violating ${\pi}N$ collisions, 
our prediction is still about one order of magnitude smaller than 
the existing estimates which were based on 
the other OZI-violating mechanisms, such as duality-violating 
mechanism~\cite{Bolzan}, a generalized Veneziano model~\cite{Kodaira}, 
and on the analogy of  the $\pi^-p{\to}{\phi}n$ 
reaction~\cite{Berger}.
 
However, we should once again recall that there are, as yet, no  
experimental data for the exclusive 
$\pi^-p{\to}{J/\Psi}n$ reaction.

\vspace{1cm}
We would like to thank U. Mosel and A.W. Thomas for a 
careful reading of the manuscript and making suggestions. 
Our thanks also go to W. Cassing for productive discussions. 
This work was supported in part by the Forschungszentrum J\"ulich, 
and the Australian Research Council.

%
\end{document}